# How well-intentioned white male physicists maintain ignorance of inequity and justify inaction.

Melissa Dancy

Western Michigan University

melissa.dancy@wmich.edu

Apriel K. Hodari, Eureka Scientific, Inc.

## Abstract

Background: We present an analysis of interviews with 27 self-identified progressive white-male physics faculty and graduate students discussing race and gender in physics. White men dominate most STEM fields and are particularly overrepresented in positions of status and influence (i.e. full professors, chairs, deans, etc.), positioning them as a potentially powerful demographic for enacting systemic reform. Despite their proclaimed outrage at and interest in addressing inequity, they frequently engage in patterns of belief, speech and (in)action that ultimately support the status quo of white male privilege in opposition to their intentions.

Results: The white male physicists we interviewed used numerous discourses which support racist and sexist norms and position them as powerless to disrupt their own privilege. We present and discuss three overarching themes, seen in our data, demonstrating how highly intelligent, well-intentioned people of privilege maintain their power and privilege despite their own intentions: 1) Denying inequity is physically near them, 2) Locating causes of inequity in large societal systems over which they have little influence and 3) Justifying inaction.

Conclusions:  Despite being progressively minded, well-meaning, and highly intelligent, these men are frequently complicit in racism and sexism in physics. We end with recommendations for helping these men to engage the power they hold to better work with women and people of color in disrupting inequity in physics.

## Key Words: Equity, gender, race, discourse analysis, whiteness, masculinity

# Introduction

> *"There's something else I was going to say... oh yeah, anger. It makes me angry to think about race and gender in physics because I think there's so much wrong and there's so little I can do about it. Honestly that's not a small part of why I don't plan on continuing in physics after grad school, is that I don't think I can have enough of an impact that I can be not just continuously furious with the culture that I'm stuck in."* - Ryan, white male physics graduate student

Ryan is not unusual. He finds himself in a field where people like him, cis white men, are granted unearned power and opportunity. He recognizes the inherent injustice of his own privilege and is angry about it. Like many white men, he did not choose this reality and desires to dismantle it. He reports, *"I am fairly strongly involved in working on the culture of physics. I'm pretty well educated on a lot of these things, more than I would say an average selection of my peers. I've been taking equity trainings since ninth grade. I've actually run a couple of equity trainings in physics. …I'm not alone. I think there are a lot of other people like me."*

We agree with Ryan, there are a lot of other men like him. Men who deeply care, men who are willing to give their time to learning and acting to address the injustice they see around them. And also like Ryan, many of these men feel powerless to have an impact.

This powerlessness is perplexing. White men dominate physics (and most STEM fields) numerically at all levels, with their overrepresentation increasing for positions of status and influence (i.e. full professors, chairs, deans, etc.). White men historically had the most influence shaping culture and structures in STEM and, despite much rhetoric and action for reform, continue to hold disproportionate influence today. They are the holders of power and yet they frequently position themselves as powerless to address inequity.

There are white men who deliberately fight to maintain their unearned privileges but they are a minority. In a 2020 survey of 1023 chemistry, math and physics faculty in the US (Dancy, 2022), the vast majority of white male STEM faculty (86%, n=440) classified efforts to encourage diversity in their field as "beneficial" while less than 2% classified such efforts as "detrimental". Likewise 91% selected agree or strongly agree to the statement "I have a personal responsibility to take action to address in equity in {my field}."

As this data demonstrates, the majority of white men who study and work in physics recognize inequity exists, desire for a change, and are willing to exert personal effort toward that end. And yet, it continues. Despite significant time and resource expenditures over the last 25 years, the percent of women and people of color earning a PhD in physics has increased only marginally (Porter, 2019). And for those with intersecting minoritized identities, representation is still close to zero. For example, out of 59,894 PhDs awarded in physics from 1972-2017 only 90 (0.15%) went to Black women (Miller, 2019). In this study we explore why it is that inequity continues when nearly everyone in the field wants change. Specifically we address the basic question, **by** **what mechanisms do well-meaning white men uphold white and male supremacy in physics?**

We posit that answering this question requires a lens that is rarely utilized, studying those who hold the most privilege and therefore the most power to shape the culture and structures that maintain inequity, i.e. white men. Yet, the majority of research and intervention projects focus on those who hold the least power. For example, efforts to increase representation typically focus on increasing interest and preparation among underrepresented groups (Hill, 2010), including support structures such as bridging programs, mentorship, and financial support (Dickens, 2021, Ashley 2017), and inclusive pedagogies (Dewsbury 2019). All of these interventions are directed at changing those who hold the least power in the system while leaving those who hold the most power out of the discussion of both the causes and the solutions to inequity.

We strongly support efforts to better encourage and support the presence and success of those who have historically been denied access to STEM. However, we are critical of the corresponding lack of efforts to understand and impact those who have historically been granted an overabundance of access.

Here we take up the question, what barriers and possibilities are presented by those who hold privilege and power in dismantling inequity in STEM? Specifically, we use a critical lens to analyze interviews with 27 self-identified white male physics faculty and graduate students, probing their knowledge, beliefs and experiences with race and gender issues in the context of physics. Through the lens of critical discourse analysis, we look not just at the conscious intention of the expression of these men, but at the impact of the words they use and the belief system underlying how they engage as white men in a racialized and sexualized society. We analyze how the words they use convey power and lead to their complicity in racism and sexism.

*A note on language and focus.*
The topic we undertake is very complex and ideas are constantly evolving. In order to focus and engage in discourse we need to simplify concepts that are not simple. We focus our study on understanding privilege as it applies in the contexts of gender and race. There are many other contexts in which the oppression/privilege dynamic is salient in both society and in physics. Individuals experience unfair access to opportunity based on many different attributes beyond race and gender such as disability, sexual orientation, gender identity and expression, age, class, religious affiliation, native language, immigration status, and physical appearance. We acknowledge there are many other identities that also impact a person's experience.

We use the term "people of color" to identify people who experience oppression based on their real or perceived race. We acknowledge there is debate about the appropriate term to use. The acceptability of terms is constantly in flux as richer understandings of racial oppression evolve. We also acknowledge gender is not binary. When we refer to men we mean people who experience greater opportunity due to their gender or gender expression, i.e. cis men.

# The possibilities and perils of white men as disruptors of sexism and racism.

There is a large body of DEI-based research focused on the experiences, beliefs, and actions of people in STEM who hold minoritized identities (e.g., Hill 2010, Blackburn 2017, Estrada 2016). In contrast, research focused on those with privileged identities is sparse. The dearth of research focused on understanding the experiences, beliefs and actions of the people who hold the most power is not mere oversight, but rather illuminates how power influences what research questions can be asked and of whom (Prescod-Weinstein, 2020, Schiebinger, 2002).

Within STEM, white able-bodied heterosexual men hold intersecting identities of extreme privilege. They benefit from the highest levels of social inclusion, professional respect, career advancement opportunities, and annual salary, while experiencing the least harassment at work (Cech, 2022).

Access to a wide range of advantages supports their persistence in STEM and results in uneven representation across demographics. Within physics, 64% of bachelor degrees (APS, 2018) are awarded to white men (81% are awarded to all races of men and 79% are awarded to white students). The disparity only increases with increasing positions of power. For example, 90% of full physics professors are men (Porter, 2019). Data on the number of full physics professors of color is unavailable, but anecdotally, they are sparse. By a large majority, full physics professors are white men.

# White men can confront sexism and racism with less negative personal consequences and are listened to more than women and people of color.

Because of the overrepresentation, privilege and access to positions of power white men in physics hold, they are afforded unique opportunities to impact systems and culture around them. When people of color and women try to confront racism and sexism they frequently experience numerous negative consequences while the racism and sexism they were confronting goes unaddressed. For example, of discrimination complaints filed with the Equal Employment Opportunity Comission (EEOC) between 2010 and 2017, 82% of complaints received no form of relief (Jameel, 2019). Data from 2012-2016 indicates that the most likely outcome (63%) of filing a complaint with the EEOC is for the target of the discrimination to lose their job (Tomaskovic-Devey, 2021).

Members of minoritized groups who give voice to the sexism and racism they experience and witness on campus are likely to experience numerous negative consequences. For example, Kaiser & Miller (2001) found that an African American student complaining about racial discrimination was viewed as a complainer and evaluated less favorably even when there was an objective likelihood that the discrimination had occurred. Women and people of color who give voice to their oppression are viewed as complainers and troublemakers, told they are overreacting, experience having their values dismissed and become targets of retaliation (Gulker, 2013, Kaiser, 2003, Czopp, 2003).

People from privileged groups have very different experiences when confronting oppression. They experience few negative consequences and are frequently rewarded for their efforts, even when small (Drury, 2014). For example, in an interview study, Patton & Bondi (2015) found that white male faculty reported experiencing few negative outcomes as a result of their ally work, while generally being rewarded on campus by recognition for their efforts. Similarly, Czopp & Monteith (2003) found that privileged individuals were able to confront bias while eliciting less irritation and antagonism than their oppressed counterparts. Eliezer & Major (2012) found that when men confront sexism they are not viewed as complainers as much as women who engage in the exact same actions.

Importantly, when someone from a privileged group confronts racism or sexism they are much more likely to be listened to and to have the issue they are confronting addressed. For example, Czopp & Monteith (2001) found that when a confronter was part of a privileged group they elicited more guilt and corrective responses from the aggressor.  Similarly, Rasinski and Czopp (2010) found that when a white person confronted racist behavior they were viewed as more persuasive, and the aggressor viewed as more biased, than when the same behavior was confronted by a Black person. And Drury (2013) found that targets are taken less seriously, are less believable and are judged to be more overreactive than privileged members when confronting prejudice.

Because of their status, white men have the privilege of being able to stand up to racism and sexism without experiencing the negative impacts that people of color and women face for the exact some confrontation. They may even experience positive impacts such as professional recognition for their ally work. And because they are listened to more, the result of their effort is more likely to see the racism and sexism addressed. It is therefore essential for white men (and all people in privileged positions) to take an active role in dismantling the unjust, oppressive structures they themselves benefit from. The cost and risk for them is less and the rewards greater.

# White men's ignorance leads to their complicity in racism and sexism even when they are well meaning.

White men are essential for dismantling sexism and racism. However, their privileged status, while conferring on them the ability to be listened to more and punished less for speaking up, has not helped them to recognize and respond to oppression around them. This ignorance results in their support for the status quo even as their intention is to undermine it.

The first step in confronting oppression is to recognize it exists. In this regard white men are challenged. For example, Dancy et. al. (2020) found that white male undergraduate STEM majors were far more likely than other undergraduate STEM majors to declare that race and gender have no impact on a person's experience as a STEM major. When white men did recognize race and gender have an impact, they rarely attributed it to unequal opportunity, preferring explanations related to innate differences and choices made by members of different groups. Others report similar findings (Swim, 2001, Rodin 1990, Becker 2011). Once oppression is recognized, the next step is developing motivation to act. In this step white men are also challenged as they frequently minimize racism and sexism as something not serious enough to warrant addressing. Bonilla-Silva (2018) identifies the minimization of racism as a common framing of race-evasiveness.

The next step after recognizing oppression exists and warrants action is to actually act. Research on allyship indicates that while people in privileged positions have a lot to offer and can be impactful in their action, they too frequently undermine the goals of their own intentions and end up doing harm. The ally is in a challenging position working to dismantle systems they have benefited from, while embedded in a culture that encourages those with privilege not to see or understand that privilege. In order to not do harm, the ally must devote considerable effort to understanding privilege and their complicity in it. Without such work they can easily fall into problematic patterns (Reason 2005). For example, those from dominant groups often interact with their oppressed counterparts from a paternalistic point of view, positing themselves as a savior rather than partner (Trepagnier, 2017, Endres 2009) and they too frequently focus their efforts on superficial vs. structural change (Edwards, 2006, Reason 2005). They can also easily fall into performative allyship (Kalina, 2020), for example, posting outrage on social media but never engaging deeply with the issues.

People who hold identities that are conferred unearned privilege (i.e. white, male, able-bodied, cis gender, upper-class, heterosexual, etc.) have tendencies to use discourses that support the status quo of power. Sometimes these discourses are intentional, such as those used by white supremacists. However, much of it is done without intent or awareness of those who engage in it.

For example, Bonilla-Silvia (2018) documents the pervasive ideology of color-blind racism. Statements such as "I don't see race" or "there is only one race, the human race" can appear superficially as race neutral. However, they have the impact of negating the real experiences of people of color and therefore have the impact of reinforcing white supremacy.

Many others have written about other moves progressives make that obscure oppression and support the status quo. Much of this comes out of the literature focused on race. For example Sue (2015) writes about the moves white people make to avoid acknowledging race including: claiming society is meritocratic, positioning racism as a thing of the past, minimizing differences or pretending not to see them (color blindness), refusing to see power and privilege, and denying their individual role in racism. Likewise, in a study of white male undergraduates, Cabrera (2019) found the common moves of attributing segregation to natural tendances instead of racism, locating racism geographically separated from themselves, viewing racism as a thing of the past, and using "yes, but" statements to acknowledge and then minimize racism.

Dancy, M. & Hodari, A. "How well-intentioned white male physicists ..."

In our initial analysis we started with many of the discourses identified by Pleasants' (2011) study of men taking a women's study class. The men in Pleasants' study, like ours, were progressive and committed to equity but engaged in a number of discourses that ultimately supported their own privilege. Identified discourses included: Appeals to self (turning the focus of conversation to their own reactions and intentions), appeals to progress (minimizing inequity as something that is getting better or inevitable), and appeals to authority (discounting evidence of oppression by dismissing the methods of data collection).

DiAngelo (2021) has also written extensively about the discourses of well meaning white people that perpetuate racism. These discourses, most of which we saw in our interview data, include discourses related to establishing oneself as a good person who is not racist, downplaying racism and one's own advantages, justifications to deny the existence of racism, expecting BIPOC people to teach white people about racism and justifying not taking claims of racism seriously. Many others have identified similar patterns (Daniels 2021, Matias, 2016, Finders 2020, Hytten 2003, Knowles 2014).

All of these discourses serve to maintain an appearance of being anti-oppression but actually serve to support the privilege of the speaker by obscuring the existence of and real life impacts of oppression. People of privileged identities are enculturated into a system that teaches them to use these problematic discourses and ultimately maintain the status quo even when their intentions are to do otherwise. In this study, we document common moves of progressive white men in physics and the impact that has on their ability to impact the status quo they claim to want to disrupt.

## Summary and Guiding Questions

The research record clearly indicates that people who hold privileged identities are essential to disrupting inequality. They are able to speak up and work for change without the risks and with the benefits of being listened to and taken more seriously than those who hold oppressed identities. Notably, white men are able to have a greater impact, not because they hold more knowledge or ideas, but because other white men, who hold most of the positions of power, listen to white men more than other groups. Sexism and racism ironically leads to white men having an ability to be heard as they confront sexism and racism.

While white men have an extra ability to confront sexism and racism, their good intentions are insufficient. Having an impact requires them to extend significant effort to understand the role they currently play in maintaining sexism and racism. It will also require them to work along with people of color and women and not recenter their own experiences. It is important that white men not take charge, but rather support the work of others.

Although many people of privilege are well-intentioned and are motivated to work for change, they frequently lack an understanding of inequity and their own complicity in maintaining it. We therefore strive to illuminate the ways this lack of understanding shows up, in order to provide insights into ways to disrupt it. Specifically we use interview data and critical discourse analysis to explore an important question.

*By what mechanisms do well-meaning white men uphold white and male supremacy?*

# Methodology

## Interview Participants

We utilize interviews with 27 self-identified white male physicists talking about race and gender. Interview participants were identified by advertising through our personal networks. A description of the study, identifying its focus on race and gender, was circulated to people who had access to appropriate networks (i.e. to share on a faculty or graduate student email list, or send personally to colleagues and/or acquaintances). We primarily targeted people who were at institutions with a large physics graduate student population. The study notice contained a link to a survey for those who were interested in participating to share basic demographic and contact information. In order to participate in our study, interviewees had to self-identify as white and male and be either a graduate student or a faculty member in a physics department at an academic institution in the United States. Men who responded to our survey and fit the basic study criteria, were then contacted by someone on the interview team to schedule an interview which was conducted in person or virtually depending on locations and preferences.

Because discussing race and gender is frequently sensitive and uncomfortable, interviews were conducted by one of four white men on the research team in order to match demographics with the participants. The interview team was not involved in the development of the protocol or in the analysis of data. The interview team all had a STEM background and previous experience in education research and in qualitative methods. Interviews were audio-recorded and professionally transcribed for analysis. Interviews took place over the course of several months in 2019.

|  | N |
|---|---|
| Graduate Student | 14 |
| Faculty<br>    Untenured Faculty (n=2)<br>    Tenured Faculty (n=11)<br>    Current or Past Department Chair (n=4) | 13 |
| Doctoral High Research Institution | 23 |
| Other Institution | 4 |

*Table 1 - Status and Institution Type of the 27 Interviewees*

We were able to interview 27 male faculty and graduate students from 13 institutions across the United States. Most of our participants were at research universities, including all of the graduate students. Table 1 displays the number of participants by status and type of institution.

Toward the end of the interview we asked participants if they had any other identities they felt impacted their experience in physics. Table 2 details the number of participants identifying each identity. Some participants gave more than one salient identity. Since this question was open-ended, the exclusion of a participant from a particular identity does not mean they did not have it, just that they chose not to mention it as salient.

| Salient Identity | Number of Participants |
| --- | --- |
| Jewish | 12 |
| None Identified | 5 |
| 1st Generation College | 3 |
| Non-Native English Speaker | 3 |
| Middle Eastern Descent | 2 |
| ADHD | 1 |
| Neuro Diverse | 1 |
| Low SES Background | 1 |
| Catholic | 1 |

*Table 2 - Other identities participants identified that are salient to their experience in physics. Question was asked open-ended so respondents could give more than one answer and the lack of an answer does not indicate the identity is not present.*

Our interview sample is not representative of the overall population of physicists. We recruited through informal networks and participants were self-selected. This most likely resulted in a sample that is more concerned about issues of inequity in physics than the overall population. All of the men we interviewed saw themselves as generally knowledgeable and concerned about equity issues. They frequently reported being involved in equity related work. As graduate student Dan articulated during the interview *"I just want to contribute to whatever this study's goals are. How can I be an effective ally, you know? I go to protests, try to have the right state of mind, counter my own oppressive thoughts and behaviors that I know are there because I was raised as a white male. And what else can I do? Well, maybe I can contribute to a study that'll help try and directly solve the problem."* Dan's expression was not unusual for the men who volunteered to participate in this study.

Because of the way we sampled, our findings are specific to progressive white men. As detailed in the literature review section above, this specific group represents an important demographic for understanding how to disrupt sexism and racism, as they are the ones most likely to advocate for change and to be listened to by other white men. Efforts focused on helping progressive white men diminish their complicity in racism and sexism and in supporting them as agents of change are essential.

Dancy, M. & Hodari, A. "How well-intentioned white male physicists ..."

All participant names are pseudonyms, most were chosen by the participants themselves. In some cases participants chose names that were overly distracting (such as Pokemon or science fiction characters), prompting us to rename them so as not to draw unnecessary attention to the names.

## Protocol Development

The interview protocol was designed to probe for participant's understandings and experiences with race and gender in a physics context and was designed in collaboration between the two authors. The full protocol can be found in Appendix A. We utilized two types of questions as described below.

Four questions presented participants with data from other studies and probed for their reaction to and explanation of the findings. Three of these presented quantitative data. We presented participants with the most recent data on representation in STEM, data from a pew research center study reporting on levels of discrimination in STEM (Funk, 2018), and results from a study demonstrating bias in hiring in STEM (Moss-Racusin, 2012). We also shared data from a qualitative study demonstrating the isolation of a high achieving Black female physics student (Johnson, 2007). In each case participants were asked for their initial reaction with follow up prompts the interviewer could use to deepen understanding of their reaction and ensure salient aspects were discussed.

Additionally, other questions were used to more directly probe participants' ideas. Examples of questions in this category include "Does sexism or racism exist in your department?", "Have you ever witnessed discrimination?", "Tell me about a time you discussed race or gender over the last year.", and "Would your advice for someone interested in pursuing physics depend on that persons race or gender?"

The interview protocol was designed to be semi-structured. Interviewers were encouraged to follow ideas brought up by the participants and to keep the interview conversational. Each question had follow up prompts to be used, depending on the answer given by the participant, to ensure salient aspects of the question topic were discussed if not naturally brought up.

The initial version of the protocol was piloted by the authors with three men similar to the ones who participated in the study. Small revisions were made based on feedback from these men. The protocol was also used by the authors to interview the four men who conducted the study interviews, providing another source of validation and feedback as well as ensuring interviewees understood the deeper background of each question so they could better direct follow-up questions.

Dancy, M. & Hodari, A. "How well-intentioned white male physicists ..."

# Analysis Methods

Each interview was audio recorded and transcribed for analysis in the qualitative software program Nvivo. Some codes were developed a priori based on our understanding of important topics of note and areas identified in the research literature that were likely salient. We were particularly inspired by the work of Pleasants (2011) in our a priori codes. The coding strategy we used was largely the inductive approach found in grounded theory (Glaser & Strauss, 1967)

We used the lens of critical discourse analysis (Fairclough 2013, Mullet 2018, Wodak 2015) in order to make sense of and draw conclusions from the interviews we conducted. Critical discourse analysis draws attention to the discourses used, either intentionally or unintentionally, to illuminate and understand social inequities. Through this framing we focused on the words our participants used and asked how those words, in their collective meaning, either support or undermine power structures related to race and gender. Critical discourse analysis views the impact of what is said relative to social relations as meaningful outside of what is intended. It is therefore a way to understand dynamics of oppression which would be consciously denied by the speaker.

The critical discourse analysis lens is helpful here because all of the men we interviewed were progressively minded and not intentionally engaging in sexist or racist behavior. It is only through a critical lens that the relationship of their patterns of belief and action to power can be seen. While these men all hold "well-meaning" beliefs, they are embedded in a culture that strongly maintains inequity and encourages them to participate in this maintenance. Through the lens of critical discourse analysis we can begin to identify and name the mechanisms by which they hold and use their power.

Each of the authors read an individual interview and made notes of statements that indicated something about the participant's beliefs and actions around equity, with a critical eye toward what these statements indicate about how they support or disrupt social inequity and power. We then met to go over each interview and talk through commonalities and discrepancies in our individual analysis. With each interview, codes were developed describing aspects of their discourse. Over time these codes were expanded, refined, and collapsed. The process was then repeated until all 27 interviews had been analyzed and discussed with agreement reached by both authors. Each interview was given extensive attention by both researchers.

# Positionality Statement

This work presented here is inextricably influenced by the author's identities and experiences. Dancy identifies as a middle-class white cis-women. Her professional experiences are largely centered in academia, teaching, and educational research. Throughout her career she has sought ways to promote equity and contribute to the body of knowledge on racism and sexism. Additionally, she has studied and worked extensively in physics departments similar to those of our participants and has extensive experience engaging white male physicists in conversation and reform initiatives related to race and gender. She has many frustrating experiences engaging white men in dialog which impact her work on this project. As a woman in physics, she has experienced many instances of limited opportunity due to her gender. As a white person, she has experience with and identifies with many of the problematic perspectives of privileged white men.

Hodari grounds her positionality in her identities as both a small Black girl who fell in love with mathematics at age seven, and a working-class city kid whose hungry brain found a home in our shared discipline. Her career has centered the lived experiences of successful women of color in STEM education and careers, set in non-academic research and policy spaces, as well as K-16 educational institutions. The experience of engaging the knowledges and beliefs raised in the study underlying this paper has underscored Hodari's positionality relative to the interview participants' privileged gender and race. This experience has also highlighted the racialized power systems we all live within, and how these systems project onto our lives the complexity that comes with living in different worlds, despite the commonality of our disciplinary home.

## Epistemology of Ignorance

We utilize the theoretical framing of the epistemology of ignorance to make sense of our interview data. Epistemology is concerned with the mechanisms of how we come to know. The epistemology of ignorance (Sullivan & Tuana, 2007) concerns how someone can not know that which is right in front of them. How is it that someone can fail to know something that is obvious? As Sullivan and Tuana (2007) articulate

> *"Ignorance is often thought of as a gap in knowledge, as an epistemic oversight that easily could be remedied once it has been noticed. It can seem to be an accidental by-product of the limited time and resources that human beings have to investigate and understand their world. While this type of ignorance does exist, it is not the only kind. Sometimes what we do not know is not a mere gap in knowledge, the accidental result of an epistemological oversight. Especially in the case of racial oppression, a lack of knowledge or an unlearning of something previously known often is actively produced for the purposes of domination and exploitation … it can take the form of the center's own ignorance of injustice, cruelty, and suffering." (pg. 1)*

We make sense of findings using the epistemology of ignorance perspective. This perspective posits that ignorance about racism and sexism is not a matter of lacking the opportunity to know, but rather that ignorance serves the interests of those in privileged positions (i.e. white men) and is therefore cultivated and maintained by systems of power, both from individuals and also from structures. In other words, ignorance serves as a tool to maintain white and male supremacy.

Using the epistemology of ignorance framing we approach our data with the following questions. What do these men not know? How does their not knowing support white and male dominance? And more specifically, what are the mechanisms by which these men maintain their ignorance of racism and sexism in spite of excessive information all around them that both exist in close proximity to them and significantly impact their non-white male students and colleagues. In other words, we focus our analysis and interpretation on understanding exactly how it is these otherwise highly intelligent and successful physicists maintain high levels of ignorance of sexism, racism and their own privilege. Identifying and understanding the mechanisms of ignorance offers clues to mechanisms for dismantling it.

# Findings

We find that these men, despite professing pro-equity beliefs, pervasively engage in patterns of thought, speech and action that work to uphold white and male supremacy in physics while justifying their own personal inaction. Below we detail three patterns that illuminate how these progressively minded men maintain their ignorance of racism and sexism and justify taking little action to disrupt it.

All themes discussed were commonly found in the interviews of both faculty and graduate students. We find no support for the oft-asserted idea that the problems are largely located in the older generation.

## Theme 1 - Physical Distancing: Inequity happens in places far far away.

Our interviewees all indicated they believe inequity exists in STEM. However, they generally view inequity as something that does not happen anywhere physically near them. They commonly talk about it as located at a physical distance, where they have little influence. From their perspective, it is something that exists, but it is not happening in their classrooms, in their research group, in their departments, among their colleagues, in their geographic area, or in their field. By denying the inequity near them, they maintain ignorance of sexism and racism while positioning themselves as unable to take direct action.

### It doesn't happen in my classes.

When presented with evidence of inequity in physics, participants frequently responded by indicating that while there may be inequity in physics it is happening physically elsewhere. For example, senior faculty David, after reading about the common experience of social isolation of Black women in STEM courses (Allen, 2022, Johnson 2007), explicitly denied it would happen in his own classroom stating,

> *"I did not encounter something so explicit and so in your face like that. I do have not many, but a few African American or people of color in my classes, and I never encountered that. The first thing in class that I say to students is I discuss these kinds of behavior so they understand very well. There's no way to misunderstand the message that I have. So perhaps in my classes it doesn't happen because of that. I don't know."*

We saw no indications in David's response that he was earnestly engaging in understanding and considering the information he was just presented. Instead we see David defending what was happening in his classroom and justifying his classes were different because he *talks* to his students, a strategy which is unlikely to result in meaningful impact. He also makes no mention of using feedback from minoritized students to assess his classroom environment, positioning his own perceptions as accurate, without effort to understand the perspective of others.

### It doesn't happen in my department or research group.

Denial that inequity is present in their department, though they agree it happens in other departments was very common. For example, when asked what advice he would give a potential student of a minoritized faculty member Brian stated he would tell them,

> *"My department tends to be fairly male-dominated. … who you're going to be interacting with is mostly white men,… I would say race and gender isn't something that's part of the equation in people's interactions. I think that the people that I work with, they don't discriminate on the basis of race or gender. … I have had conversations like that with students. Like, just be aware, you will be accepted, you will be welcomed, you will be valued. No one will think for a second that you are less than in any way."*

Brian acknowledges his department is largely white and male and he gave no indications he had authentically engaged women or people of color in his department in discussions of departmental climate. Yet, he is confident his department is welcoming to people not like him.

Similarly, senior faculty Mark, after a discussion about problematic student-student interactions indicated *"It's not something I see day to day…I think the microaggressions in this environment are less explicit."* And when asked if there was sexism or racism in his department, graduate student Steve responded, *"In my department, if it does that would be especially embarrassing given {the work the department has done to address equity} … if we do have a problem with sexism that's really kind of embarrassing."* Both of these men are in denial about the inequities that almost certainly exist in their own departments.

A discourse move similar to denying issues in their department was to acknowledge their department may have issues but claim their research group was immune as graduate student John explained *"I'd like to think it doesn't really exist within {my research area} group. That's evidenced by the fact that we have open discussions just amongst the graduate students about these issues."*

### It doesn't happen in my institution.

A particularly common physical distancing move was to locate inequity outside their own institution. For example, graduate student Dane after being presented with data illuminating the isolation Black women frequently encounter in physics classes stated he could see it happening because *"there are places in the United States where racism is still rather prevalent, even though people are uncomfortable talking about it. So I imagine there are some locations in the country where this could certainly happen."* Graduate student Leon responding to the same prompt likewise indicated that *"If I saw this on the news or something I wouldn't be surprised."*

Dancy, M. & Hodari, A. "How well-intentioned white male physicists ..."

And faculty member Paul similarly stated *"I mean it doesn't surprise me that somewhere in the US that would be the case. …I think {at my institution} at least I haven't seen anything like that. People are much more integrated."*

When participants claimed discrimination was at other institutions but not their own, they sometimes justified this claim by explaining that at their institution people were good-hearted as graduate student Francis explained, *"I don't think I've seen {discrimination}. I can't have a particular experience in mind. Because I think most people at {my university} are pretty sensible about how to behave."* Likewise graduate student Jason claimed his school would not tolerate such behavior *"I went to a fairly liberal undergrad where you probably would've gotten doxxed as a racist or something if people found out that this was happening."*

Notably, when a specific location of inequity was mentioned it was overwhelmingly in the south, a region where none of our interviewees were located. Faculty member David talked of a man of color who told him of his experience with discrimination in physics noting *"he is from South Carolina. No, from Georgia, he was from Georgia."* Graduate student Joe acknowledged discrimination happens but not *"in the context of physics students, more or less. I have seen stuff like this happen at {a university in South Carolina}."* And faculty member Paul stated *"There have been all these stories about a faculty member in North Carolina."*

### It doesn't happen in my field of physics.

Participants commonly distanced themselves from inequity by distancing it from their academic field of physics. For example, denying the evidence of sexism and racism in STEM by claiming that it was being driven by fields other than physics, because physicists are more progressive, or more thoughtful than other scientists. This despite physics being one of the least diverse of STEM fields.

For example, Joe, a graduate student, in considering the impacts of race and gender in physics stated, *"That's probably one of the advantages of physics is you get to work for smart people, so normally {race and gender is} not an issue."* When presented with data demonstrating race and sex-based discrimination in the STEM workplace, he attributed it to non-physicists, *"Well, if you're going to be working in an industry you're going to eventually wind up working with people who are not STEM. ….So you do get some of that boy's club."* He likewise explained away any impacts in physics classes by attributing racism to non-majors taking physics classes, *"Fortunately in physics nowadays they've basically quarantined physicists into major classes. I can see {racist behavior happen} if you get non-majors thrown in there."*

Similarly, AJ, a senior faculty member, after being presented with the same data demonstrating workplace discrimination in STEM, attributed it to engineers, *"{The data is from} all of STEM, so that includes engineering. Attitudes among engineers are a lot different than attitudes among physicists … in my experience 90 to 95 percent of all physicists view themselves as liberal …. but engineers that's not the case."* And faculty member Scott similarly explained that the incident of discrimination he had just been presented from a physics classes was *"not an example of racism in science or physics just because you've described this large lecture class,*

Dancy, M. & Hodari, A. "How well-intentioned white male physicists ..."

*so I assume this is a freshman or sophomore level class. It's mostly not people who I would think of as physicists. Some small fraction of them might be physics majors. So that's more an example of racism in society."* Senior Faculty Mark similarly blamed data demonstrating discrimination in STEM on those in "*computer jobs*" where *"there are some pretty serious bro type cultures out there in the work environment."*

The high levels of discomfort these men feel in acknowledging inequity exists physically near them is evident in the way they minimize it when they can't discount it directly. Many of our interviewees who could not deny it exists near them made a point of immediately following their admission by stating that it isn't just physics. This frequently came up when we asked participants to comment on the existence of racism and sexism in physics specifically. Examples of this distancing move include: *"I guess I'd have to say yes. Just to be clear, I say that because I believe that it's everywhere"*, Scott (senior faculty), *"It's not just physics. I mean this is happening all over the place...I don't doubt there's sexism in physics, but there's sexism other places too."* Jonah, (senior faculty), *"I think race and gender impacts people in every area."* Jay, (senior faculty), and *"I think it was wider than just physics, I think it was all sciences."* Mark (senior faculty). Although these participants were able to acknowledge that sexism and racism are in physics they responded to this direct question with "Yes, but" thereby distancing sexism and racism from their own field to place it in the domain of everywhere.

## Summary of Theme One: Physical Distancing

A common discourse move was to locate inequity at a physical distance. While all participants acknowledged inequity happens in physics, they also talked about it in terms that locate it far away from themselves where they would have little or no influence. They commonly claim that it isn't an issue in their classrooms, their research group, their department, their geographic region or in their specific field, or subfield. When justifying their claims they frequently appeal to the goodness of the people around them or the efforts made in their department in addressing inequity as evidence that it does not exist. Notably, their claims are not backed up by any evidence nor do they generally report having exerted effort talking to and listening to women or people of color in their department to inform their opinions of the superiority of their local environment.

The framing of inequity as something that is located physically out of reach is quite problematic as it positions both the causes and solutions of sexism and racism in physics as outside the sphere of influence of these men. If they cannot acknowledge that inequity is something that exists in proximity to them, they can not even begin to use their privileged influence to dismantle it. This distancing move is also a mechanism by which these men maintain their overall ignorance. If it isn't near them, there is nothing to be known.

# Theme 2 - It's too big for me to impact: Locating inequity in grand societal structures.

As described above, physically distancing themselves from inequity was a common way participants located inequity outside their sphere of influence and therefore justify not acting. Another common way they did this was to attribute the causes of inequity to large societal structures over which no one could easily have influence, particularly the participants themselves. By viewing causes of inequity in grand societal structures, they can absolve themselves of the personal responsibility to act. As graduate student John articulated when he posited the solution to inequity was *"big societal shifts, right? We need to raise kids of any gender without imposing the cultural expectations of what that gender does."* John went on to acknowledge the helplessness of this view of the root of inequity stating *"I don't have a single concrete suggestion for how to do that. Because you can't just change society."*

In our interviews, we saw our participants commonly invoke many grand societal structures as they explained the source of inequity *in physics*. Here we summarize four common structures identified by the participants: The K12 education system, socioeconomic factors, societal expectations of parenting responsibilities, and historical legacies of overt sexism and racism.

## It is the fault of the K12 educational system.

A common way our interviewees explained the cause of inequity in higher education physics was to blame levels of the educational system below the one in which they teach and operate. By locating the cause of inequity in what came before, they can free themselves of responsibility to act in their local context.

Below we summarize comments that focus on gender. In the next section, we take up comments focused on racial inequity because their explanations for the failure of the K12 system around race are conflated with ideas of socio economics.

In explaining why women are underrepresented in physics, faculty member David stated, *"I think that the biggest bottleneck in our field is high school."* He went on to blame high school teachers for discouraging women and attributed it to them not being trained in physics *"it's a message that we as a society pass to women, that they are not welcome here…that's a message high school teachers pass to students. I think that the worst offenders are high school in general. …most of the teachers that teach physics in this country never got a degree in physics, they don't really know physics, they fear physics."*

Likewise, graduate student Joe offered that women are not pursuing physics because *"I think they're herded out before it comes to that graduate or undergraduate level."* And faculty member Larry *"you have to put a lot of money into making high schools in particular more equitably funded and more sensitive to gender and race issues than they are currently."* And faculty member Scott *"It's clear that it is an imbalance that starts to show up in high school. … whatever it is, it's happening at a relatively young age."*

While it is well documented that there is attrition at levels before higher education it is likely all of these men are in departments in which women and people of color disproportionately leave the physics program compared to their white male counterparts. Yet, they focus their attention on causes that occur at levels below them over which they have almost no influence. By attributing the cause of the overrepresentation of white men in physics to the K12 system they relieve themselves of engaging in learning and action in their own departments.

## Lack of racial diversity is explained by socioeconomic dynamics.

Overwhelmingly, the most common explanations for why physics is dominated by white people centered on unequal opportunities before arriving at the university, either because they view schools with students of color as being inferior or because they believe students of color are poor and therefore without financial means to pursue physics.

### Students of color attend schools that provide inferior preparation.

There was a pervasive belief in the inferiority of schools that students who are not white attend. For example, faculty Mark in explaining why whites are overrepresented in physics stated,

> *"I think the US educational system also can disfavor minority populations in substantial ways in terms of resources and what educational opportunities are available to students. It can bias towards and in some sense boost the white majority in that there could be just better opportunities for them where they live and in their communities and so forth. So I think that has something to do with it, because physics is one of those things that takes a lot of background, and that background has to go back years. You can't just start in college of course."*

Mark believes poor quality K12 schools where people of color live are responsible for the lack of students of color in his department. Similar comments by other interviewees include,

> *"Well, so for the under-representation of Black and Hispanic students I think a ton of that can be attributed to... places with high Black and Hispanic populations tend to have substantially worse schooling."* Graduate student John

> *"I mean we know that African Americans in this country are overall less well off than white people. This is going to show up in what opportunities do they get exposure to in school. I mean how many students are seeing a physics class in high school and getting the chance to have the sort of inspiration I did growing up?"* - Faculty member Jonah

> *"I see a particular underrepresentation of Black students. I think that probably has to do with preparation, schooling. I think a lot of Black students, certainly the African American students that we get at my university, tend to come from communities where the schooling isn't as good. They tend to be somewhat less prepared for college."* - Faculty member Brian

Dancy, M. & Hodari, A. "How well-intentioned white male physicists ..."

These men are framing the racial overrepresentation of white people in physics as a problem of K12 schools by asserting that students of color are provided with an inferior K12 experience, leading to their under-preparation which leads to them not pursuing physics. While it is well documented there is discrimination in education that negatively impacts non-white students, the argument that this accounts for the wide disparity in racial representation *in physics* is severely flawed. There are academic fields for example, in which students of color are not underrepresented, despite enduring a discriminatory K12 experience. It also does not account for the higher attrition rate of students of color over white students once they have entered into physics. Additionally, these men are using deficit model framing and exhibiting racial bias by claiming the majority of students of color are underprepared. By focusing on problems in the K12 system our interviewees maintain their ignorance of the racism and sexism in their local environment and justify not taking action.

Poverty explains the overrepresentation of white people.

In a similar line of reasoning, our participants also commonly attributed disparities in representation to people who are not white coming from poverty. This often came in the form of arguing that one needs economic resources to afford to pursue physics. As graduate student Alex explained, *"There is a large financial burden to get into grad school. That's just what you have to get through undergrad, so any sort of disparity there is going to carry through. …so if you have groups that are not getting through undergrad for financial reasons or other reasons then it's going to carry through to your statistics on grad school."*

Faculty member Larry similarly argued that non-white people have less economic advantage which leads to being too unprepared to pursue physics stating, *"Early preparation is really important as far as preparing for a physics career. I think that's probably true no matter what, whether you're interested in an academic career or some other type of career. But if you're not inspired early on, if you don't have a good background in mathematics before you enter college, I think it becomes ten times harder to succeed in a physics career or in any career that involves a good mathematics preparation. Statistically, I believe it's true that Hispanics and African Americans are economically disadvantaged compared to Whites."*

Several of our interviewees went even further, attributing the lack of representation to both poverty and the choices those in poverty make, namely the choice not to go to graduate school. They are blaming students of color for their own oppression. Examples of this discourse move include,

> *"Graduate school is strongly correlated to your economic background. Because to go to graduate school you have to go to undergrad, so you have to go to a good undergrad to get into graduate school. Good undergrads are expensive in the US. …When you're a minority and you're not wealthy, you go to a job. You go {to college} to get something that has a job at the end."* - Graduate student Francis

*"To get a PhD, that takes a lot of time and money … Black and Hispanic kids are just typically lower income, so if they go to school they just want to get that job and start making money. If you have student loans you've got to pay them off pretty fast. White kids, yeah, typically come from higher backgrounds."* - Graduate student Joe

*"I think there are probably lots and lots of factors. I think there's probably no single solution. In general, I believe in the US that minorities, especially Hispanics and Blacks, have a more difficult time obtaining an education due to many economic factors. I think economics can play into this a lot. Most people don't perceive getting a PhD as necessarily the easiest or cheapest thing to do, or most affordable thing, so people from lower socioeconomic statuses, which tend to be more Hispanic and more Blacks in the United States, have a harder time getting into these systems."* - Graduate student Vince

Arguing poverty prevents people from pursuing physics and people of color are poor therefore physics is overrepresented by white people is not only flawed but also quite harmful. The idea that poverty and race are so closely associated is a figment of mainstream media more than of reality. In a study of race, poverty and the media (Dixon, 2017), it was found that the media depicted poor families as Black 59% of the time even though they account for only 27% of families living below the poverty line. Additionally, recent census data (Creamer, 2020) indicate that the vast majority of Black people (80%) are not poor. Attributing poverty as the cause of the dramatic underrepresentation of Black people and other people of color in physics simply does not hold up under even basic scrutiny. And as with the other themes presented here, it diverts attention from the direct causes related to unequal opportunity in departments of physics in institutions of higher education.

### The historical legacy of overt sexism and racism are to blame.

Another grand structure participants blamed for the disparity in physics was the historical legacy of sexism and racism. The argument being that sexism and racism, things of the past, continue to exert influence because change takes time to be fully realized.

For example, graduate student Jason in explaining racial disparities in representation in STEM stated the cause was *"cultural inertia"* which he explained as *"The experiences of {different races} of people in this country have each been very different on the whole than the other groups."* And faculty member Jonah when asked if race or gender impacts someone's experience in physics responded, *"I wouldn't doubt it. I'm sure this is more difficult for women than for men because of some of the historic environment on it."* Jonah went on to describe the way women feel out of place at conferences with mostly men because the legacy of past sexism has resulted in the current overrepresentation of men. And faculty member Paul *"With African Americans we had slavery, so there's that, and then we had segregation. Then we had Brown V Board which was supposed to fix this but we know that it doesn't do anything. …it's kind of a no brainer as to why the representation hasn't been there historically…. the lack of representation I think is largely correlated with the fact that we haven't rectified the injustices that have been done to certain groups."*

Dancy, M. & Hodari, A. "How well-intentioned white male physicists ..."

In addition to expressing generalized and often vague ideas of historical inertia to explain disparities, our interviewees commonly explained it in terms of an older generation that has just not yet left the field. Not unsurprisingly, this theme was expressed almost exclusively by the younger graduate students who were themselves clearly not in the category of old white guy. Examples include,

> "*I can definitely see especially an older generation of PIs thinking that women for whatever reason are less competent at doing whatever job it is, and that probably stems from a long career of just institutionalization.*" - Graduate student Leon

> "*I think a lot probably in interactions among maybe older white male professors. They take a lot of heat, but I think there are some that don't take women as seriously as they should, that don't take minorities as seriously as they should.*" - Graduate student Vince

Graduate student Ryan went a little further claiming the older men were presenting an active barrier to change, *"I feel like there is a slow push for making the culture of physics better and more inclusive and more equitable and there is a giant wall of old white guys with tenure that are going to be here for another 20 years until they drop dead standing in the way."*

Graduate student Chuck argued that his generation has had a lot of education about bias issues so they will do better, *"I do think that my generation, especially amongst those of us who are college educated, just have had the presence of these problems drilled into their head repeatedly to the point that many are frustrated by them and kind of joke about them like they're BS even though they're kind of true. But I think nonetheless there are just so many people who have some recognition that these problems exist, I just find it hard to imagine that that's going to have zero effect.*" Likewise graduate student Francis also felt that the solution to hiring was simply to have younger people making the decisions because they are less embedded in the bias stating, *"You make sure you have young faculty who make the interview….Because the older you get in that thing, the more you self-average. The younger you are, the more open you are. The older you are, the more ground up into your workplace you are. It's like you've got to move from the 65s to the 40s."*

When blaming today's problems on the culture of yesterday, while viewing the people of today as not of that culture, the solution becomes to simply wait for the older folks to retire. Those who are younger, or older but awakened as the younger generation is, therefore do not need to take any direct action. By maintaining an ideology of sexism and racism as things of the past to be cured shortly through retirement, these men can justify to themselves not taking action to challenge and address structures of inequity.

## Differential expectations of parenting explain the gender gap.

The final grand societal structure we describe here centers on the differing societal (or personal) expectations of family responsibilities for women and men. In explaining the gender gap in representation in physics, a common explanation was either that society had different expectations for women and men as parents, which makes it more difficult for women to be in physics, or that it is women who want something different that is not compatible with being in physics.

For example, faculty member Jonah explained the gender gap in physics to the extra responsibility women have in parenting over men, *"There are certainly some societal problems of family expectations of women, also related to having children, and the extra work that accumulates for mothers that doesn't accumulate for fathers. That can certainly make things more difficult for women in the field. There certainly is a big culture of now, now, now, we have to get this done, and that doesn't always fit well with families."* And faculty member AJ gave a similar explanation stating, *"If you're a grad student for six years, so 22 to 28, you're definitely impinging on when you might want to start a family. … So I think there are challenges, and maybe they have a bigger impact on women than they do on men."*

Some interviews went a bit further. In addition to asserting society has different expectations of men and women, they also used the deficit model argument that women make different choices thereby blaming women for their own oppression. For example, faculty member Paul attributed the gender gap to women wanting to have children sooner and being more willing to give up their careers for family. *"The fact that the field is not very great at accommodating people that want to have families. There's a correlation with that, that women might be more inclined to have families at early stages and men can maybe put it off and remain a bachelor until a much later stage in life and so forth. Or also the social pressures on women are historically common, and I think persistently they're more willing to sacrifice careers than men."* Likewise, faculty member Chris in explaining the gender gap attributed it to the desires of women stating, *"I was just talking to a female colleague who was really struggling with the idea of balancing a longer career versus having a child, things like this. …- I've never had a male colleague have this discussion with me ever in my life, and I've had at least three or four female colleagues have this discussion with me I think both before, during, and after, their PhDs."*

None of the men we interviewed acknowledged or gave an explanation of how it is that women thrive in other fields while having children, yet not physics. Unequal parenting expectations do impact women who enter the workforce, but they do not explain why physics has such a large gender imbalance. As with the other identified grand societal explanations for representation gaps, attributing the gap to either the way society treats motherhood vs fatherhood or the different decisions made around parenting by men and women, absolves these men of personal responsibility to act. They can't impact societal expectations of parenting and they certainly can't impact the decisions women make. This perspective leaves unexamined what is happening in their own classrooms, research groups, and departments.

## Summary of Theme Two: Grand Societal Structures

When asked to explain inequity in physics it was common for our interviewees to invoke very broad cultural structures over which they have no control. Here we discussed four of the most common.

1. Blaming the K12 system for discouraging women and providing inadequate education to students of color.
2. Attributing the lack of racial diversity to the incorrect idea that most people of color live in poverty.
3. Arguing that sexism and racism are mostly of the past and change will occur over time without further action, especially when the current group of old white men retire.
4. Attributing disparity to unequal expectations of parenting for mothers and fathers or the different choices made by women and men regarding parenting.

While it is true that there are grand structures that negatively impact the participation of people of color and women in physics (there is a high rate of dropping out of the physics pipeline in high school for example) by invoking these structures as the main cause of inequity in physics in higher education these men absolve themselves of needing to act personally. Additionally, as we demonstrated, their ideas about the nature of the structures they evoked were often flawed and unsophisticated in their understanding (e.g. that racism and sexism are the result of legacy). All of the structures they evoked are general, and not specific to physics, a field with more significant representation problems than other fields. They make no attempt to account for why these structures, which are not physics related, can be used to explain why *physics* has significantly less people of color and women than other fields.

By focusing blame on people and structures over which they have little to no influence they maintain their own comfort as they exist in departments where people of color and women are likely to exit physics at higher rates and experience less support than their white male colleagues. By promoting grand structures as their explanation they are able to maintain their ignorance of what is happening directly around them.

## Theme 3 - I'm helpless to act, therefore my inaction is justified

> Interviewer: *"How does it feel to talk about race and gender?"*
> Graduate Student Zander: *"No particular feelings. Maybe it's creating a sense that I need to do something about it, but not enough to actually go significantly out of my way to do something. Plus I don't actually know even what going out of my way would be."*

The above exchange with Zander represents a common expression of our interviewees, though most were not as forthcoming about the unlikelihood of their action. The themes presented previously demonstrate the ways these men position themselves as helpless by locating inequity in places over which they could not possibly have any influence. In this section we consider their response when confronted with situations over which they clearly have the potential to have influence. For example, when a situation is brought to their attention happening in the same classroom or department they occupy.

Throughout the interview there were many opportunities for our interviewees to share personal experiences with inequity and discrimination. Sometimes such stories came up naturally as they discussed topics, other times it came up in response to directly being prodded to share their experiences and other times they responded to hypothetical prompts we provided. When they brought up stories on their own or responded to situations we brought into the discussion, we generally followed up by asking what they had done in response, or for the hypothetical, what they would do in such a situation. For the most part, our participants report very few attempts to intervene when inequity is happening directly around them and provide numerous justifications of their inaction.

In this section we discuss three common patterns of belief and cognitive process participants used to position themselves as helpless to act. First, they often deny even obvious oppression, freeing them of the responsibility to address that which they refuse to see. Secondly, they cite perceived negative consequences of action, to themselves or others, to justify inaction. And third, they position themselves as incapable of acting either because they lack knowledge or skills needed to act.

## Refusing to see as justification for inaction.

Before discussing their articulated justifications for not acting, it is important to note that they frequently deny much of the inequity that is happening around them, even when it is obvious. We saw evidence in our interviews of men using ignorance to justify not engaging in addressing an issue of equity. We turn again to graduate student Zander whose candor we appreciated.

> Interviewer: Have you ever witnessed discrimination?
> Zander: Probably.
> Interviewer: Probably?
> Zander: Have I experienced it as such? Probably not. Have I talked myself into believing that it wasn't really discrimination? Probably true. So I've probably witnessed discrimination and have not recognized it as such. I've probably witnessed discrimination and recognized it and convinced myself that it was not actually discrimination or that I don't need to take any action. I don't think I've ever witnessed discrimination and taken action against it.

Zander's exchange with the interviewer exemplifies a common pattern by which those with privilege justify not acting. While Zander was able to recognize that he might be encountering inequity but not allowing himself to see it, it is generally a justification that is made subconsciously. It is true that people can lack knowledge because they have not had the opportunity to know. But here we highlight cases where the lack of knowledge was not due to lacking opportunity to know, but rather a not knowing despite overwhelming exposure to knowledge, a "willful" ignorance.

A clear example can be seen in faculty member Brent's discussion of the surprise he felt at "learning" his research group of 40ish members had only one female member.

> *"One of the first graduate students I hired was female and that didn't seem very strange to me…I never even thought about it. But interestingly, speaking to her sometime later, she of course arrived in the largest research group there and immediately noticed, as I*

Dancy, M. & Hodari, A. "How well-intentioned white male physicists ..."
24

> *would have done, and I hadn't really seen it from a distance, that she was the only female other than the group secretary, in a group of five or six professors, a dozen postdocs, twenty graduate students. She admitted that if she'd known that she probably wouldn't have come." - Faculty member Brent*

It is beyond plausible that Brent did not notice the extreme gender disparity in his research group. By refusing to see, he was able to absolve himself of action to address the sexism that was rampant in his field and research group.

Another example can be seen in faculty member Paul's response to reading a first hand account of a Black woman in a physics classroom (Johnson 2007). In the reading presented to Paul, he learned about a Black woman who reported no one sat next to her in her physics class and if she sat next to other students they would move to different seats the next class period. After reading the account Paul stated, *"I mean if I were her I probably wouldn't do anything. I would just be like okay. If she's sufficiently self-aware she probably understands why that's happening, so if I were her I'd just be like okay, this is the world that we live in."*

It is dubious that anyone with the level of intelligence it takes to become a physics faculty member would think that a student in this situation would be ok being isolated by her peers. Paul continued, fully entrenched in his willful ignorance stating, *"You know, maybe she wants to sit alone. We can't just assume that she- you know, maybe she just wants to be alone."* Paul had just read an account in which this student expressed clear dismay at the situation and was clearly attempting to shift the situation by taking the initiative to join other students and yet he still convinced himself that she wanted her fellow students to move away from her. By framing this clearly racist situation as something the student would be ok with, or even want, Paul was able to justify not taking action.

Similarly, in response to the same reading, graduate student Joe was asked what he would do if he were the professor of the classes. Joe stated that he wouldn't do much because the student, *"may not want {you to address the situation}. Trust me, if they're Black they're used to it by now unfortunately...If I personally had that happen, {I would be like} just fuck them, focus on your work, I didn't come here to socialize."* It is obvious from the account Joe read that this student was experiencing racism that significantly limited her opportunity to learn and that she was not at all ok with it. Despite how obvious this was, Joe told himself that the obvious was not true. Through his cognitive clouding he is able to distance himself from the reality of the world he occupies and therefore justify not addressing the racism around him.

## Action would create negative consequences.

We now turn our attention to justifications participants gave for not acting when their proximity to a situation warrants action. A main way participants justify not acting, even when they recognize inequity is happening within their sphere of influence, is to appeal to the negative consequences of acting, either for themselves or for others.

*Negative consequences for me: I don't want to experience discomfort.*

Several of our participants expressed their reservations about acting due to their discomfort in confronting others. For example,

> "*I'm not a very confrontational person, especially in social situations, so though it might sadden me, even if I noticed it and was like oh, that sucks, I don't know that I would necessarily do anything.*" - Graduate student Leon

> "*I hope that I would have the moral courage to either say something or do something, but I'm not sure that I would. I tend to be fairly non-confrontational.*" Faculty Larry

> "*I don't want the situation to exist, but if I'm the one who solves it I'm going to have to have this discussion that's going to force me to talk about things I'm not necessarily comfortable about.*" - Faculty Brian

A number of our participants appealed to the seniority and power of the person behaving in sexist or racist ways as justification for not speaking up. The following exchange with graduate student Ryan reflects this justification for inaction.

> Interviewer: "*In these moments where you've felt like you might have seen some discrimination … have you ever responded or taken any action to what you witnessed?*"
> Ryan: "*No, I haven't. I think part of it is the power differential. In {the case we just discussed} I was a very new grad student so I felt like I had no power and that my position was potentially somewhat tenuous.*"

Graduate student Joe, in considering how he was personally impacted by sexism and racism, lamented how difficult it would be to witness sexist behavior from his supervisor while ignoring it to get through without making waves. "*It's everywhere. As I said, it'll kill morale if you know for a fact you're working for a sexist asshole, especially if you're a PhD student and you just need to finish up your dissertation. A lot of that motivation is just to keep your head down and hope you can get through it.*"

Likewise faculty member AJ, after bringing up an incident where he witnessed inappropriate comments by a male faculty member was asked by the interviewer "*With respect to these comments, the snide comments you mentioned from colleagues, how did you respond when those comments were made?*" to which he explained he did not do anything because "*Some of them were when I was a student, and I didn't engage. I'm certainly not brave enough to tell some more senior person that that comment was over the line.*" AJ attributes the power differential between him and others as justification for not intervening.

Similarly, faculty member Scott described a situation he witnessed in which he did nothing justifying his inaction because he was quick on his feet and because the person was in a powerful position. "*No, I didn't {act}. There's a couple of pieces to that. One is I've never been very good at thinking on my feet, so I'm not sure I would have been able to come up with the* *right thing to say. Another is just in that situation this is a senior faculty member at a top institution, so I probably wasn't going to challenge him. I think it was just afterwards we looked at each other and said what was that?"*

Negative consequences for the perpetrator: I don't want others to feel uncomfortable.

Several participants attributed their lack of action toward a problematic situation near them by appealing to the impact on those who were creating the racist or sexism environment. For example, graduate student Leon expressed concerns about using his position of power as an instructor to call out problematic student-student interactions explaining,

> *"You can't coerce other students to change their behavior because at that point you're now in a position of power and that's sort of- it's not really right to use your power to, I don't know, shame people into doing things, especially when that could then negatively affect your impartiality towards their grades in that course. You know, it's bad that they're treating this person awfully, but it would also be bad to punish them in that kind of way. It's sort of a weird lose-lose situation. You would just wish that the thing didn't happen but, you know, these are people in college. They're adults. They have their own things. They should be held accountable for the things that they do, but that seems like an inappropriate way to exercise that ability. You would want larger social pressures to help guide them into the correct course rather than you taking retribution on them."*

Leon's account is interesting and full of contradictions. As an instructor he presumably uses his power to get students to do things all the time, for example, he likely controls what they work on, when they speak, what they speak about, etc. It is likely if a student were talking loudly on their cell phone during a lecture and disrupting the learning of the class as a whole, he would not feel it inappropriate to ask them to stop. But when it comes to disrupting the learning of another student based on their race or gender, Leon claims you can not "coerce" students using your "power" because it would "shame" them. He then places the responsibility for acting on behavior in his own classroom back on "larger social pressures".

Similarly, faculty member Alan expressed concerns about upsetting students who lack an understanding of racial privilege, *"I'm also sensitive to the way sometimes that language can be taken as offensive to white students. I don't like to refer to white skin privilege in the context of a situation where I don't know very well the students with whom I'm speaking, because I know students who are white whose fathers have had a tough time and become alcoholic and sometimes drug addicts, and they don't feel privileged. They're not taking that as a social justice question, they're taking that as a personal affront. Nothing is gained that way."*

What is concerning about these accounts is how the comfort of the perpetrator was prioritized over protecting the target. This view was expressed by only a few participants but that it occurred at all among this group of progressively minded men is concerning, especially since we didn't explicitly elicit it.

**Negative consequences for the victim: I don't want to create further harm to those who experience sexism and racism.**

Several of our participants brought up concerns about causing further harm to the targets of racism or sexism should they try to intervene. For example, faculty member Chris discussed his hesitation intervening because *"there are things you don't know how to do correctly, that you are afraid of saying the wrong things and things like this."* Likewise after being presented with data about bias in hiring, faculty Brent explained why he was opposed to prioritizing hiring for diversity. *"We have an assistant professor search going on now. I happen to be chair of the committee. So then you ask a question about how do you make sure you do a really diverse search to make sure you get the people there. …But that provokes a very complicated discussion about how you get there. And of course one of the things that I think you can't do is to make gender and racially profiled hiring, not even because it's illegal but because frankly it wouldn't serve the people you hire. If somebody can whisper, well, the only reason she got the job was because she's female and we had a better candidate and we couldn't get them past the team it would cause issues for everybody."*

A common microaggression is to assume that anyone hired, who is not a white man, was only hired due to their minoritized status even when their qualifications exceed the other candidates. The backlash Brent refers to is likely to happen simply because a woman or person of color was hired. Further, it is notable that Brent's comment immediately followed being presented with data indicating that men and women with identical resumes would not be evaluated equally when being considered for a STEM job. Brent did not offer any ideas for how to negate the bias that is known to exist except a vague *"whenever we're grading applicants we have a rubric, a set of questions."* Without interviewer prompting, he then turned the conversation away from bias to blaming the choices women make for gender disparity stating *"I think it's associated with the fact that there are women who are making choices, and they have to decide am I going to become a theoretical condensed matter physicist where I'd be seeing a whole bunch of gray white males or go over here to something where I see a different demographic."* Collectively, all of these moves by Brent allowed him to not directly consider how he, as the chair of a hiring committee would intervene in the gender discrimination he knows is likely to occur.

Participants frequently brought up potential negative consequences to the target when considering how they would intervene as an instructor upon witnessing problematic interactions between students. For example,

> *"Obviously you can't {call out bad student behavior} because that would be an excruciating experience for the person who {is the target}….. even if the rest of the class might pick up on it and internalize that, they're a little bit ashamed that that happened and that they played a role in that happening, there's n people who get to absorb that together versus the one person who has to deal with the fact that they're being presented as, no matter how it's phrased, the cause of this problem, that if they just weren't there this wouldn't have been noticed."* - Graduate student Chuck

> *"I honestly don't know how to effectively counter it without causing even a bigger backlash…There could be active animosity towards the student."* - Faculty member Jay
>
> *"There's not much you can do to change people's minds that's not going to cause a scene, disrupt the class, or other stuff. {The target} may not want that."* - Graduate student Joe
>
> *"I probably wouldn't do much…. and maybe if she hadn't noticed, all of a sudden I pointed it out to her and made her feel awful?"* - Graduate student John

These men are expressing hesitancy to address problematic racialized behavior in their own classrooms out of worry there will be a backlash from the white students who are creating a hostile learning environment for a student of color. They justify not acting as being what the target of the racist behavior would want, when in reality it is the discomfort of these men that is actually underlying their inaction.

## My inaction is justified because I am not capable of acting.

Finally, we turn our attention to moments when participants acknowledged seeing inequity, acknowledged something should be done, but positioned themselves as incapable of doing anything. For example, graduate student Ryan indicated he would not address a situation in his classroom because *"I feel like I don't have the social skills necessary to tell a class don't be doing that."* Presumably, Ryan, as the lead of his classroom, has many experiences telling students what they should be doing. It is only when it comes to guiding the students around equity related behavior that he tells himself he is unskilled. Below we provide further examples of how these men justified their inaction based on their perceived inability to engage.

### I'm not capable because I don't know what to do.

A common reason given for not acting was a lack of knowledge about what to do. For example,

> *"I'm not great about stepping into that conversation and being productive about this. My sense is oh, that's awful, but then I don't do much with it. So part of it is I don't feel like I have - and maybe I do - but I don't feel like I have the tools to handle this productively."* - Faculty Mark
>
> *"I'm not sure I would even know how to try to make a situation like that better."* - Faculty Larry
>
> *"{If someone asked me for advice about an equity issue I} would not be able to give useful advice. …I mean I could offer sympathy, but that's probably not what they're looking for in that context."* - Graduate student Steve

Frequently, our interviewees followed their assertion that they would not know what to do with a claim that if they found themselves observing a problematic situation they would need to reach

Dancy, M. & Hodari, A. "How well-intentioned white male physicists ..."

out for advice. For example, graduate student Jason, after being asked what he would do if he were the instructor in a class where an issue arose responded, *"I don't know. I would probably have to ask for advice from somebody else, ask my friends. I would probably reach out to people I know and be like what do you think I should do? Should I do anything? Should I only give advice? Something like that. I don't entirely know what I would do."* Likewise, faculty member Paul stated, *"I would have to consult somebody else and think about it through time and just be like I have to get back to you, I don't know what the solution is."* And faculty Jay, *"I don't know that I would know what to do. But I know that there is a faculty member in the provost's office and one of her specialties is dealing with issues of diversity and inclusion. I would go and talk to her to get advice on what to do."*

We note that although it was common to suggest they would go to someone else for advice in dealing with a problematic situation, none of our participants indicated they had actually done this in a real life situation. We question if they would actually seek out help. It is likely they simply would not act, justifying their inaction to themselves because they didn't know what to do.

I'm not capable, it is someone else's responsibility.

It was common for our participants to go further by positioning themselves not just as lacking knowledge about how to act but as unable to know. A good example is seen in faculty member AJ's response to how he would deal with issues in his own classroom. *"I have to say that this kind of thing is not my strong suit exactly. …I might actually refer {the target of racialized behavior} to someone else... Our university has enormous- you know, it's a private school, lots of support staff for students. Dean of students, dean of freshmen. There's lots out there. I might actually refer them to someone who's a little bit more professional. I'm not a very good therapist really."* Although we were asking about how he would manage his own students in his own classroom, AJ deferred to others on campus and ended with a claim that he was not skilled enough. It is also very problematic that AJ equates dealing with race based hostility in his own classroom with being a therapist. He views taking action in this case as someone else's responsibility.

Similarly graduate student Dane also said he would refer students encountering racism to others on campus stating, *"I would try and get them in touch with whatever organization on campus specializes in working with these kinds of students to help connect them with resources and whatnot. … I know there are a few groups or organizations that specifically try and assist minority groups on campus to make them feel more welcome and whatnot."*

We also frequently saw our participants directly or subtly claiming their status as a white man prevented them from knowing and acting. As faculty member Chris articulated, *"I continue to not have great advice … part of it is just that my experience has been so different that I do struggle with trying to provide good advice in these types of situations."* Likewise when asked if sexism and racism existed in his department faculty member Jonah replied "*Good question. I would like to think no, but am I in a position to guarantee that? You'd have to ask some of the women.*" In this case Jonah is positioning himself to be unable to know about the situation in his own department because of his gender.

Dancy, M. & Hodari, A. "How well-intentioned white male physicists ..."

Graduate student Dan' explanation of why he feels he has more understanding around gender than race is informative.

> *"I can compare my trajectory in my field to my girlfriend's. …I don't have many people of Hispanic or Black race to compare that kind of thing to. I mean I have Indian colleagues, but they tend to almost share a similar position of privilege as whites {provides an example of a highly successful fellow Indian graduate student) ….But I don't know if it would go similarly for someone who might be Black or Hispanic. I simply don't know. I just don't have that data. I don't have that comparative experience because they're just not there for the most part. And the few that are there,  there's one in our department …, he's a great guy, but he's not in my same field so I don't see him on a daily basis. I never see him in a professional context and I'm not close with him, so we never get to talking about the day to day professional experience of doing physics and then I never get to compare that kind of thing. That's I think why that blind spot is there, is I don't have a lot of contact with people of that origin and therefore I simply have a hard time comparing."*

Although Dan has a race and a gender, both of which impact his experiences regularly in life and the physics context, he believes himself unable to understand racial or gendered experiences without someone from the marginalized group to "compare with".

I'm not capable because no one is, it is impossible to address.

Another way we saw our participants justifying not acting was to appeal to the ultimate level of helplessness, that there is nothing anyone could do, positioning sexism and racism as inevitable. We classify this under not knowing what to do because it is never the case that racist or sexist behavior can not be addressed. The belief that racism and sexism can not be addressed is at the extreme end of claiming inaction is justified due to lack of knowledge.

The impossibility of addressing sexism and racism showed up in two ways. One was to claim that it was impossible to not be biased therefore sexism and racism will always exist. Zandar, Dan, and Chris all spoke to this.

> *"There is sexism and racism, and we will never get rid of it, but I don't think it's intentional … you will always have a bias. You will always be biased by your experience and most likely your identity in one way or another."* Graduate Student Zandar

> *"You're asking me what we should change here? The change is make people not biased, but that's sort of an impossible thing to do."* Graduate Student Dan

> *"I think it's nearly impossible for me to remove that bias….It's something that I struggle with, and I think that we as a department always struggle with any time we do a hire."* Faculty Chris

Another way the belief in the impossibility to impact showed up was to claim others' behavior cannot be changed. Graduate student Joe spoke to this generally, *"As a student there's not much you can really do. As near as I can see you're going to cause a scene that's going to cause people to be embarrassed and you're not really going to change too many people's thinking."* While Graduate student Aaron, provides an example of this move in talking about addressing behavior of professors, *"I mean it's difficult. Being a tenured professor means that you don't have to say I'm sorry to anyone, pretty much, kind of, to a degree. So if someone's really stuck in their ways then we can't make them change."*

Likewise Ryan and Larry spoke of the perceived impossibility of impacting student behavior in their own classrooms.

> *"There's not really an easy solution. You can't be like no, you people can't move {away from the minoritized student}, you have to sit here. It sucks."* Graduate Student Ryan

> *"On the other hand, if you place a student into a group, how much they interact with that group is also something that's largely uncontrollable."* Faculty member Larry

By positioning the problem as impossible to solve, these men are able to justify their lack of action while maintaining their self view as one of good and moral. Once they place racism and sexism in the bin of inevitable, they absolve themselves of responsibility to act.

## Summary of Theme 3 - I'm helpless to act, therefore my inaction is justified.

We have detailed how these men justify their inaction using a variety of discourse moves. Each of these moves serve to justify their inaction while they continue to benefit from sexism and racism. Specifically we see them using three main lines of justification for inaction.

1. Inaction is justified when sexism and racism are not noticed, even when it is clearly happening in front of them and/or others are clearly giving voice to the racism and sexism around them. They do not have a responsibility to acknowledge it.
2. Acting would create negative consequences worse than the racism and sexism itself. Inaction is justified when acting would be uncomfortable for the individual or for the person exhibiting the racist or sexist behavior. Also, women and people of color often don't want racism or sexism addressed because addressing it would be uncomfortable for them or would create a backlash.
3. Inaction is justified when one doesn't know what to do. White men can not understand racism and sexism, they must depend on others to tell them what to do, or to take action for them. Also, sometimes there is nothing that can be done. You can't change other people and bias will always exist. Sexism and racism are an inevitable part of life.

Each of these moves allow our participants to maintain their image of themselves as good men, while doing little to undermine the system of white male supremacy they benefit from. By positioning themselves as justified in their helplessness they move on in their careers without using the power conferred to them by their privileged status to upend the unfair system they themselves say they want to dismantle.

# Conclusion

# Summary of Findings

Below we summarize the identified discourse moves white progressive male physicists in this study used to position themselves as helpless and justify inaction in addressing racism and sexism in physics.

**Physical distancing:** Inequity happens in places far away from me where I don't have any influence.  The people around me are good people, who are smart and thoughtful and well meaning. The people around me have had a lot of discussions about equity and we are not the problem.

- Not in my classroom: My students are good people and I tell my students not to be racist or sexist.
- Not in my research group. Everyone in my research group is a good person. We talk about inequity and how to address it so we don't have any problems with racism or sexism.
- Not in my department or university. Our department has a lot of resources and we have done a lot of work to eliminate sexism and racism. We are one of the best departments.
- Not in my field of physics. Physicists are smarter and more liberal than other fields, the bad rap STEM gets around equity is due to people in STEM from other fields like engineering.
- Not in my geographic region. Sexism and racism are things that happen in other parts of the US, i.e., the south.

*Why this is a problem: All available data indicate sexism and racism in physics are pervasive and happen in every institution. None of our male respondents offered any actual data from their own institution indicating their departments are actually different. By framing it as something physically distant, they absolve themselves of responsibility to act locally.*

**Grand societal structures cause inequity in physics.** Inequity in physics is caused by big structures in all of society over which I could not possibly have any impact.

- Gender bias in K12. Teachers and counselors discourage girls from pursuing physics and perpetuate gender stereotypes.
- Racial bias in K12. Non-white kids go to schools that provide them an inadequate education. The result is kids of color are unprepared to pursue physics.
- People of color are poor and make bad decisions. They want to get a job as soon as possible, they choose careers that pay a lot, and they can't afford graduate school.
- Historical racism and sexism explains it all.  There used to be sexism and racism which has lingering impacts even though racism and sexism no longer exist. For example, women and people of color feel uncomfortable due to being a minority which is due to historical forces not today's reality.  Today's problems are because of all the old white men who need to retire. The younger generation is aware and smart and not like the old white men.
- <u>Parenting responsibilities.</u> Society has different expectations of parenting for men and women. Also women chose to be more involved in their children's care than men choose. This explains the gender gap in physics.

*Why this is a problem: Each of these explanations fail to hold up when scrutinized. For example, not all people of color are poor, physics is a lucrative field and would be appealing to someone wanting to earn a high salary, and graduate school in physics is typically free making it a good choice for someone of low income. They all use deficit model thinking to explain gaps, i.e. women and people of color are too ill prepared to succeed either in their cognitive abilities or their attitudes and choices. Each explanation absolves the speaker of confronting what is happening directly around them. Also, all of them are general and not specific to physics yet they are used to explain why physics has greater disparity than other fields.*

**My inaction is justified.** I am not obligated to act when racism and sexism happen near me.

- <u>My ignorance justifies inaction.</u> I have trouble acknowledging even obvious sexism and racism. I can't act on what I can't see. When people tell me it is happening I don't hear them. I don't have to do anything because I don't know anything.
- <u>My discomfort justifies inaction.</u> Speaking up is uncomfortable. I don't like being confrontational. If I am uncomfortable I am not obligated to act.
- <u>Other's discomfort justifies inaction.</u> Those who perpetuate sexism and racism would feel bad if I brought attention to it by saying something to them. I shouldn't act if it means someone would feel uncomfortable being confronted with the impact of their actions.
- <u>The targets of racism and sexism don't want it addressed.</u> If I tried to take action the target of the sexism or racism I am a witness to would not like it. They don't want attention brought to their situation. It is worse for the target for me to intervene than the sexism and racism they are experiencing.
- <u>I am not capable, I don't know what to do.</u> I can't do anything because I don't know what to do. Someone else would need to tell me what to do.
- <u>I'm not capable, someone else needs to act.</u> As a white man I can't know anything about race or gender. I don't have the social skills to navigate speaking up. Someone else needs to take this responsibility. Someone from the diversity office on campus can intervene instead of me.
- <u>I'm not capable, no one is.</u> The problem is too difficult, there is nothing that can be done to address it. People, including me, can't be changed. Therefore I should not try. Racism and sexism are inevitable.

*Why this is a problem: White men, by virtue of their privilege, have the most power to intervene with the least risk of personal harm. Yet, they have many ways to justify not acting when the opportunity arises. They position their own comfort as more important than providing equal opportunity to others and tell themselves they are incapable of making a difference, even when they are in power positions and have tangible influence. Although they frequently report lack of knowledge as justification for lack of action, they rarely speak of taking initiative to increase their knowledge so they can act. Positioning themselves as helpless to act allows them to maintain their sense of being a good person while benefiting from an unfair system without challenging that system.*

Dancy, M. & Hodari, A. "How well-intentioned white male physicists ..."

# Discussion

We have presented numerous patterns of belief and action used by white progressive male physicists that have the impact of maintaining white and male privilege in physics. Collectively, these patterns serve to maintain in these men an ignorance of the sexism and racism around them, ignorance of their complicity in maintaining systems of oppression and their own privilege, and prevent them from engaging productively in reforming the landscape of physics toward justice and fairness.

We see many of the ideas underlying the epistemology of ignorance playing out. Framing injustice as something happening physically far away allows them to maintain an ignorance of that which is happening near them. Framing injustice as caused by big societal structures allows them to maintain an ignorance of the way their own field is complicit in white and male supremacy. Framing injustice as something over which they can not have influence allows them to maintain an ignorance of their own role in perpetuating it, allowing them to continue benefiting from the injustice while maintaining their vision of themselves as good, moral men.

Their misconstrued ideas allow them to maintain their ignorance and this ignorance in turn solidifies their misconstrued ideas. If they don't allow themselves to see the inequity around them, they can continue to posit that it isn't happening in their classrooms, department, or field. Their ignorance allows them to continue locating inequity far away or in grand structures they can not impact.

The epistemology of ignorance framework demonstrates how the ignorance of these men is far from accidental, nor a simple artifact of not having access to correct knowledge. Each of them are highly intelligent, well-meaning men who have had ample opportunity to engage in discussions of race and gender and to witness it themselves. Rather than being an oversight, their ignorance is developed by and maintains the status quo. A status quo that tells them it is ok to discount evidence of inequity and base assessments on their own perceptions instead of those of people who experience oppression.

It is important to not frame the ignorance demonstrated here as stupidity or immorality on the part of individual men, but rather, men responding to the culture around them, a culture which is built on a history of white male supremacy and its maintenance. We argue they have been taught not to know and are encouraged not to undo their ignorance. It will require intentional effort on their part to circumnavigate the oppressive system they are equally as embedded in as those who are oppressed.

In addition to what they said, it is also important to note what they did not say. First, no one talked about experiencing any accountability for their level of knowledge, understanding or action around sexism or racism. When they spoke of efforts to improve their knowledge or to act, it was always in the context of something they were doing out of personal altruism rather than something the structures and cultures around them were encouraging. Secondly, most of the knowledge these men had was based on their own observations. Rare was a mention of using data, or the perspectives of minoritized people, in determining the situation in their local environments. The lack of data allowed them to continue in their ignorance. The lack of accountability and a culture of allowing the perspectives of white men to define local success both contribute to ignorance and the maintenance of the status quo.

The work we report here explains how inequity persists even when everyone involved is well meaning. We see how people of privileged identities engage in a cycle of building and maintaining ignorance which allows them to maintain their view of themselves as good people while they are actually supporting oppressive systems.

# Recommendations

We offer the following recommendations based on this work.

1. **Make the target of change people of privilege.** Equity interventions in higher education STEM should target change at those who hold the most powerful positions. Support programs for those who are minoritized are positive but the primary focus should be supporting change of those who are privileged and the structures that maintain that privilege.
2. **Teach people of privilege about common discourse moves that make them complicit in oppression.** People of privilege need support to be able to understand and recognize their patterns of thought and action that lead them to be complicit in inequity. The introduction of the term "microaggression" (Sue, 2010) helped people to be able to recognize and name subtle and unintended slights. Likewise naming and promoting knowledge of the discourse moves presented here can help people recognize and decrease the use of these problematic expressions.
3. **Hold people of privilege accountable for their ignorance.** No interviewee mentioned any mechanism of accountability for recognizing or understanding equity from their departments. All of our participants were able to exist in and succeed in their field without having their ignorance significantly challenged. As long as reward structures allow white men to remain and advance in the field, while remaining ignorant of even obvious sexism and racism and prioritizing their slight discomfort over the opportunity of others to exist in the field, inequity will remain.
4. **Make equity work the work of white men.** As these men unintentionally articulated in their interviewees, equity work is frequently seen as the domain of those who are oppressed. In reality, it is those with privilege who have the power to make changes. White men need to be viewed as the ones primarily responsible for equity work.
5. **Collect, and make public, data measuring the extent of inequity in local environments.** Ignorance is maintained when perception counts as truth. It is important for departments to maintain a culture of using data to inform perceptions of how well they are doing. This includes both quantitative data as well as data obtained from listening to and believing those who are oppressed. It is important for departments to go beyond simply counting members by their demographic to gather data to understand the climate and culture in the department.
6. **Explicitly teach skills associated with confronting oppression.** Our participants frequently felt they lacked knowledge or skills needed to confront inequity. Learning to recognize and speak up effectively is a skill that requires effort to learn.

Ending racism requires white people to change. Ending sexism requires men to change. Those who are oppressed are unable to fix inequity. They can give voice to it, but they alone can not end these oppressive systems. The progressive white men we interviewed are essential to change. They have the interest and motivation to act for change and the power to have an impact. Yet they maintain significant ignorance of inequity and view themselves as helpless, despite occupying the most powerful positions. The research presented here has identified some of the ways that these progressive white men retain their self image as equity champions while actually maintaining the status quo. These results can be used to support well-meaning white men to seriously engage in the learning they need to replace their ignorance with understanding. This will allow them to engage productivity in their classrooms, departments and universities to provide opportunity for all to be in and thrive in these spaces.

# Declarations

Availability of data and materials: The datasets generated and/or analyzed during the current study are not publicly available due confidentiality restrictions but are available from the corresponding author on reasonable request.

Competing interests: The authors declare that they have no competing interests.

Funding: This study was funded by the National Science Foundation (NSF-1712436).

Authors' contributions: The study was designed by MD. Both authors jointly developed the interview protocol and managed the data collection process. Initial analysis was completed by both authors. MD completed the secondary analysis and prepared this manuscript.

Acknowledgements: We owe much gratitude to the four men who conducted these interviews, Peter Kindfield, Tim McCaskey, Joe Olsen and Seth Rosenberg. Their encouragement of respondents to fully express themselves provided a strong data set to work with. Amy Robinson and Charles Henderson provided extensive feedback on a draft of this paper which greatly improved it. We also want to thank the participants who were willing to engage bravely and honestly so we could learn from them.

Dancy, M. & Hodari, A. "How well-intentioned white male physicists ..."

Rodin, M. J., Price, J. M., Bryson, J. B., & Sanchez, F. J. (1990). Asymmetry in prejudice attribution. Journal of Experimental Social Psychology, 26, 481– 504. doi:10.1016/0022-1031(90)90052-N.

Schiebinger, L. (2002). Has Feminism changed science?. History and Philosophy of the Life Sciences, 24(3/4), 545-545.

Sue, D. W. (2010). *Microaggressions in everyday life: Race, gender, and sexual orientation*. John Wiley & Sons.

Sue, D. W. (2016). Race talk and the conspiracy of silence: Understanding and facilitating difficult dialogues on race. John Wiley & Sons.

Sullivan, S., & Tuana, N. (Eds.). (2007). *Race and epistemologies of ignorance*. Suny Press.

Swim, J. K., Hyers, L. L., Cohen, L. L., & Ferguson, M. J. (2001). Everyday sexism: Evidence for its incidence, nature, and psychological impact from three daily diary studies. Journal of Social Issues, 57, 31– 53. doi:10.1111/0022-4537.00200.

Tomaskovic-Devey, D., & McCann, C. (2021). Employment Discrimination Charge Rates: Variation and Sources. Socius, 7, 23780231211064389.

Trepagnier, B. (2017). Silent racism: How well-meaning white people perpetuate the racial divide. Routledge.

Wodak, R., & Meyer, M. (Eds.). (2015). Methods of critical discourse studies. Sage.# Appendix A - Full Interview Protocol

After going over the consent form and getting the interview participant to sign, as well as recording their code name (any first name that is not their first name), tell them you are starting the audio recording.

Start the recorder, and make some initial small talk, including introducing yourself and the participant (using codename) for the audio, along with the date and time.

Ask the participant to tell you a little about themself.

If it doesn't come up in the introductory conversation, ask why they chose physics as their discipline.

1. Tell me a brief story about your everyday life as a physicist.

2. If a younger relative or a close friend's child asked you what your social life is like as a physicist, what would you tell them?
    a. Would you recommend they choose physics as a major/discipline?  The same college/university you attended/work?

Dancy, M. & Hodari, A. "How well-intentioned white male physicists ..."

b. Would your advice depend on the other person's race? Their gender? Any other aspect of their identity?
c. If it doesn't come up … do they think people's experience in physics is impacted by gender and/or race?

I have a story from research data about an experience some students have while studying physics. I'm going to let you read it and ask you some questions afterwards. (Johnson, 2007)

One day, I walked into a huge lecture hall and saw, down at the front, one of my informants, Zina, a tall, dark-skinned African American woman. She was sitting in an aisle seat; the rest of the row she sat in was empty. I sat through class with her, and at the end of class she told me that whatever row she sits in, she clears it out—no one will sit within five or six seats of her. She explained that she used to sit in the sixth row, all by herself. Recently she had moved up to the fourth row, which had previously had habitual occupants. Now, as I saw for myself when I looked around, the sixth row held a number of students and the fourth row was empty.

I asked other African American students whether this happened to them. One told me an interesting story. She said that her roommate, also African American, said to her one day "let's go down to the library and clear out a table." She was puzzled, but they went together and sat down at a table in the library where several other students were working. Within a few minutes, all of them had left. From then on, my informant told me, she started to notice that whenever she sat down at a table, although no one appeared to notice her, within 15 minutes she was always the only person at the table even if all the other tables were crowded.

3. What strikes you first about this story? (If they don't address whether they believe this story, probe using questions like, "Does this story seem plausible to you?")

4. I am going to ask you to imagine this scenario playing out in your department. Consider what you might do if you were in various roles in relation to this story. (Use their own language as much as possible.) If needed, use some of these prompts:
    - If an undergraduate you know from a course you taught was friends with this student and wanted to help her, what advice would you give?
    - If a peer of yours saw this happening in their class, as instructor, what would you tell them to do?
    - If you witnessed this first-hand?

5. If the respondent said they'd sit next to the student, tell them that didn't happen for this student, despite the fact that many people give this response. Probe (using their language as much as possible) for why this might be.
6. (If they don't address this anywhere, ask how they think this feels to the student who experienced it.)

7. What does it mean for someone to be good at physics? Do you think your perspective is shared by your peers?

Dancy, M. & Hodari, A. "How well-intentioned white male physicists ..."

8. What is your best guess at the percentage of physics PhDs currently are awarded to men?
    a. (if they are far off) Would it surprise you to know that in 2016 82% of PhDs in physics went to men?
    b. What do you think explains this difference between men and women? Ask them to say more if needed.

9. What is your best guess at the percentage of physics PhDs among US citizens awarded to each racial category: white, Black, Asian, Hispanic? How do these percentages compare to their percentages in the US population?
    a. (if they are far off) Are you surprised to know that the actual percentages are white 87%, Asian 6%, Hispanic 4%, Black 2%, compared to general US populations of white 61%, Asian 5%, Hispanic 18%, and Black 13%?
    b. What do you think explains this gap in representation between the races? Get them to elaborate.

In a recent survey people working in STEM were asked if they have experienced discrimination based on their gender.

10. What percent of women do you think reported being discriminated against? What percent of men would you guess reported being discriminated against?

Show them the data from PEW research center, 2017.

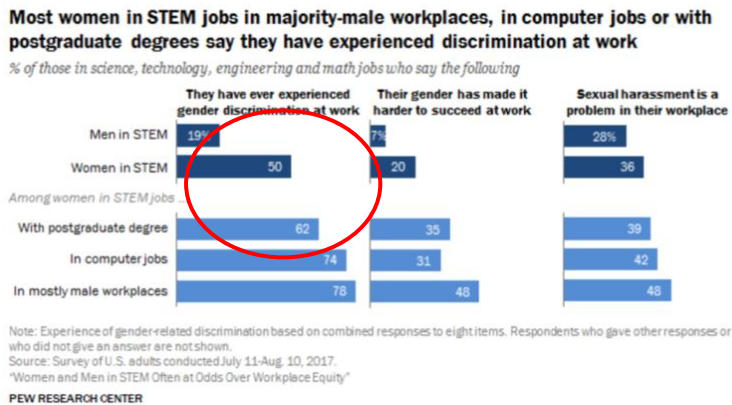

11. What comes up for you seeing this data?

12. What would you guess are the most common experiences women and men are thinking about when they report being discriminated against. (Make sure they talk about men and women.)

13. What percent of each racial category in STEM would you guess report being discriminated against at work based on their race?

Dancy, M. & Hodari, A. "How well-intentioned white male physicists ..."

Show them the data from PEW research center 2017.

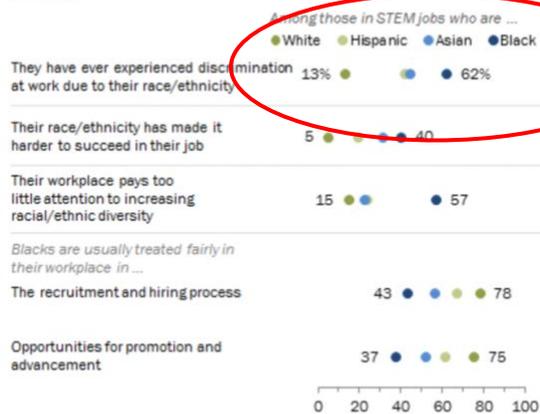

14. Does this data surprise you? What comes up for you when you see this data?

15. What do you think are the most common experiences people of different races are thinking about when they report race based discrimination?

Here's a research abstract from a recent article (Moss-Racusin, 2012). Please read it:
*In a randomized double-blind study (n = 127), science faculty from research-intensive universities rated the application materials of a student—who was randomly assigned either a male or female name—for a laboratory manager position. Faculty participants rated the male applicant as significantly more competent and hireable than the (identical) female applicant. These participants also selected a higher starting salary and offered more career mentoring to the male applicant. The gender of the faculty participants did not affect responses, such that female and male faculty were equally likely to exhibit bias against the female student. Mediation analyses indicated that the female student was less likely to be hired because she was viewed as less competent.*

16. What is your reaction to this study? What comes up for you when you hear this?

17. What are the implications of this result?

18. (If not already addressed):
    - Are these results consistent with your experience?

Dancy, M. & Hodari, A. "How well-intentioned white male physicists ..."

- Should your department make any changes based on this data? If so, what should they do?
    - (if they say a blinded process should be used) What about times when evaluations can't be blinded such as decisions about promotion and tenure, in person interviews, or daily interactions?
- Is this an example of racism or sexism in physics?

19. Have you ever been discriminated against based on your race or gender?

20. Have you ever witnessed discrimination?
    a. If no, or little, tell me more about why you have not seen discrimination.
    b. If yes, how did you react? (Probe them for any action they took and what resulted, or why they did not take action)

21. How do you identify by race and gender?

22. Are there any other identities you have that you think impact your experience in physics? If yes, follow up to understand what and how.

23. Let's revisit a question from early in the interview. Have you changed your mind at all about whether race or gender impact people's experiences in physics? If so, how?
    a. (if not addressed) What do they think the impacts are? Are impacts only for people of color and/or women?

24. Tell me about a time you discussed race and gender over the last year? With who? What was discussed?

25. How does it feel to discuss race and gender issues?
    a. Probe for both the context of this interview and also in other contexts. (if they say they have discomfort, probe for the source of the discomfort).

26. Does sexism and racism exist in physics? In your department? Get clarity on their response(s).
    a. If yes, many efforts have been undertaken to address sexism and racism in physics but the problem still remains. What could be done differently?
    b. (if not addressed):
        - Who do they think should do something?
        - On what level are changes needed (personal, departmental, global)?

27. Is there anything else you'd like to tell me about that I didn't ask?